\documentclass[traditabstract]{aa} 
\usepackage{graphicx}
\usepackage{deluxetable}
\usepackage{longtable}
\usepackage[varg]{txfonts}
\usepackage[authoryear]{natbib}
\begin{document}

\title{Mass accretion rates from multi-band photometry in the
Carina Nebula: the case of Trumpler\,14\footnote{Based on observations made with ESO Telescopes at the 
La Silla Paranal Observatory under programme ID 069.C-0426(C) and 078.D-0371(A).}}


\author{G. Beccari\inst{1}
\and
G.DeMarchi\inst{2}
\and
N.Panagia\inst{3,4,5}
\and
E.Valenti\inst{6}
\and
G.Carraro\inst{1}
\and
M.Romaniello\inst{6}
\and
M.Zoccali\inst{7,8}
\and
C.Weidner\inst{9,10}
}

\institute{European Southern Observatory, Av. Alonso de Cordova, 3107, 19001 Casilla, Santiago, Chile\\
\email{gbeccari@eso.org}
\and
ESA, Space Science Department, Keplerlaan 1, 2200 AG Noordwijk, The Netherlands
\and
Space Telescope Science Institute, Baltimore, MD 21218, USA
\and
INAF-NA, Osservatorio Astronomico di Capodimonte, Salita Moiariello 16, I-80131 Napoli, Italy
\and
Supernova Ltd, OYV \#131, Northsound Rd., Virgin Gorda VG1150, Virgin Islands, UK
\and
European Southern Observatory, Karl--Schwarzschild-Strasse 2,
85748  Garching bei M\"unchen, Germany
\and
Instituto de Astrof'sica, Pontificia Universidad Cat—lica de Chile, Av. Vicu–a Mackenna 4860, 782-0436, Macul, Santiago, Chile
\and
Millennium Institute of Astrophysics, Av. Vicu–a Mackenna 4860, 782-0436, Macul, Santiago, Chile
\and
Instituto de Astrofsica de Canarias, Calle Va Lactea s/n, E-38205 La Laguna, Tenerife, Spain
\and
Dept. Astrofisica, Universidad de La Laguna (ULL), E-38206 La Laguna, Tenerife, Spain}


 
\abstract{
We present a study of  the mass accretion rates of pre-Main Sequence (PMS) stars in the
cluster Trumpler 14 (Tr\,14) in the Carina Nebula.
Using optical multi-band photometry
we were able to identify 356 PMS stars showing H$\alpha$
excess emission with equivalent width $EW(H\alpha)>20$\,\AA. We
interpret this observational feature as indication that these objects are still actively accreting gas from 
their  circumstellar medium. From a comparison of the HR diagram with PMS evolutionary models we derive ages and masses of the 
PMS stars. We find that most of the PMS objects are younger than 10\,Myr with a median age of $\sim 3$\,Myr. Surprisingly, 
we also find that $\sim 20\,\%$ of the mass accreting objects are older than 10\,Myr. 
For each PMS star in Trumpler 14 we determine the mass accretion rate ($\dot{M}_{acc}$) and discuss 
its dependence on mass and age. We finally combine the optical photometry with near-IR observations
to build the spectral energy distribution
(SED) for each PMS star in Tr\,14. The analysis of
the SEDs suggests  the presence of transitional discs in which a large amount of gas is still
present and sustains accretion onto the PMS object at ages older than 10\,Myr. Our results,
discussed in light of recent recent discoveries with Herschel of transitional discs containing a
massive gas component around the relatively old PSM stars TW\,Hydrae, 49\,Ceti, and HD\,95086,
support a new scenario  in which old and evolved debris discs still host a significant amount of
gas.

 }{}{}{}{}

\maketitle
%

\section{Introduction}
Mass accretion is a phenomenon driving the evolution of several astrophysical objects.  It is well
known that low mass stars in the  pre-main sequence (PMS) stage grow up in mass through mass
accretion from a  circum-stellar disc~\citep[e.g.][]{ly74,ap89,ber89}.  Moreover, the process of
accretion is the power source driving the luminosities for a wide range of binary  systems
containing accreting white dwarfs and neutron stars like cataclysmic variables and millisecond
pulsars, black holes (BH),  gamma-ray bursts and supermassive BHs in active galactic
nuclei~\citep[see][]{fe09}. 
In an accreting  star this process leaves characteristic signatures in its spectrum 
mostly due to $H^+$ recombinations. Thus, one expects  
an excess in the ultra-violet (UV)  continuum, compared to stars of 
equal spectral type, ~\citep[e.g.][]{gu98,ri11}, as well as  
a strong emission in recombination lines  (e.g. H$\alpha$, Pa$\beta$ and Br$\gamma$) from the
gas ionized in the shock.

\citet[][hereafter DM10]{de10} have  presented a method  
to reliably identify PMS objects actively undergoing mass accretion in star forming
regions  using photometry. Briefly, the method combines V
and I  broad-band photometry with narrow-band H$\alpha$  imaging to
identify all stars with excess H$\alpha$  emission and to measure 
their associated H$\alpha$ emission equivalent width $EW(H\alpha)$, the H$\alpha$ 
luminosity  and the mass accretion rate, 
regardless of their age. The ability of this method to measure accurate
EW(H$\alpha$) has been discussed thoroughly and justified by DM10 and verified observationally for stars towards IC1396   
by~\citet[][]{ba11}. In their Figure\,5 these authors show an excellent  
correlation between the photometric and spectroscopic measures of EW(H$\alpha$) for emission
stronger than 10$\AA$. The DM10 method was recently applied to study PMS objects in a number of
star-burst clusters, namely NGC~3603 in the Galaxy~\citep{be10}, NGC~346 and NGC~602 in the  Small
Magellanic Clouds \citep[see][]{de11a,de13a}, 30~Doradus and three active star forming regions in
the Large Magellanic Clouds \citep[see][]{de11b, spe12}.

In this paper we apply the same method to study the PMS stars in the 
young star-burst cluster Trumpler 14 (Tr\,14) in the Carina Nebula.
Tr\,14  is a young cluster, about $1-3$\,Myr old, \citep[][hereafter
H12]{hu12}, 
whose properties have been
thoroughly addressed in the literature. The presence of variable extinction across
the cluster, as well as hints of a substantial PMS population were
initially reported by~\citet[][]{va96}, and later confirmed by Tapia 
et al. (2003)~\citet[][]{sa10}; H12. Tr\,14 is relatively close to the 
Sun, at a distance of $2.5$\,kpc~\citep[][]{ca04},
and therefore is an ideal candidate to study in detail the early stages
of stellar evolution and mass accretion.

\begin{figure*} 
\centering 
\includegraphics[width=0.9\textwidth]{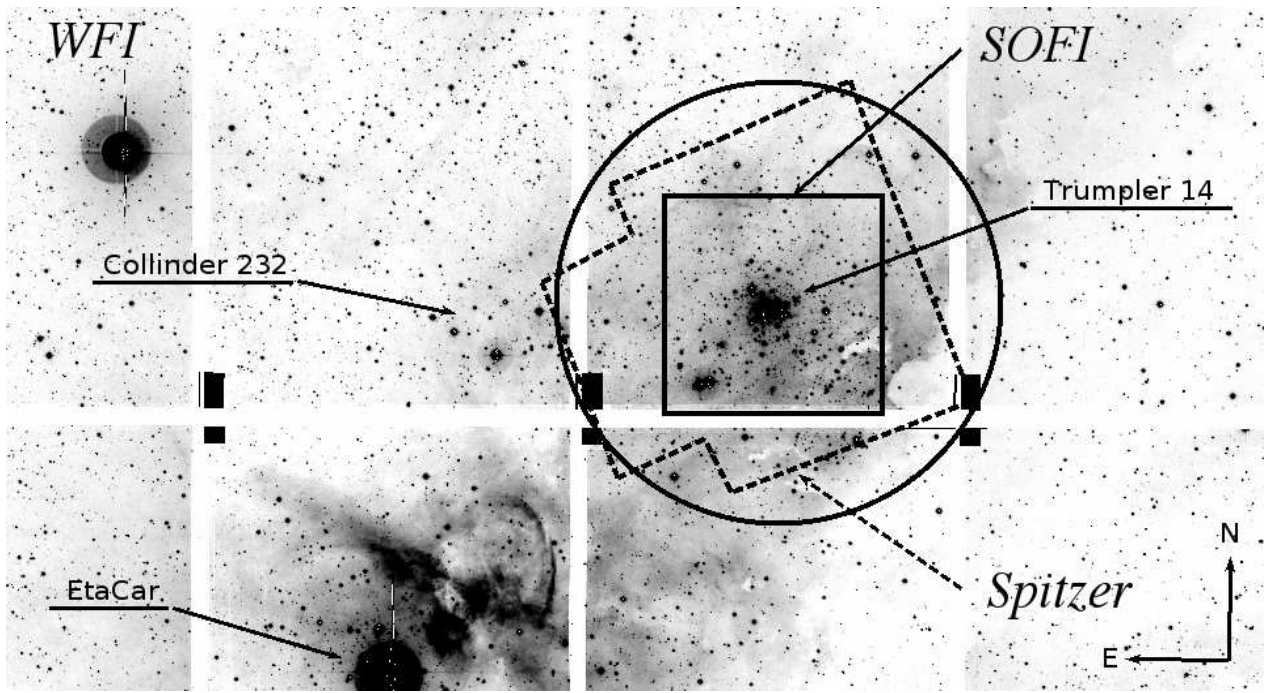}
\caption{Zoomed $27\arcmin\times15\arcmin$ mosaic image obtained with the I band filter on the WFI used to sample Trumpler\,14. 
The circle shows the area of $5\arcmin$ radius discussed in the text. The dashed and solid polygons show the area covered 
by the Spitzer and SOFI data, respectively. The location of Trumpler 14 and Collinder 232 are shown together with
Eta Carinae. The gaps between the chips of the WFI mosaic are also visible.}
\label{fig_fov} 
\end{figure*}

\section{Observations and data reduction}
\label{sec_obs}

In this work we have used panchromatic archival observations from three
different facilities, namely {\em i)} the {\em Wide Field Imager} (WFI)
at the  MPG/ESO $2.2$\,m telescope at La Silla (Chile); {\em ii)} the
near-infrared (near-IR) SOFI camera at the ESO {\em New Technology
Telescope} (NTT); and iii) the Infrared Array Camera (IRAC) on the
{\em Spitzer Space Telescope}. The details of the observations and
of their analysis are provided in Sections 2.1, 2.2 and 2.3. 
We derived in this way three photometric catalogues, all calibrated and
homogeneously registered to the same absolute astrometric system. In
Figure\,\ref{fig_fov} we show a schematic view of the combined data-set
centered on Tr\,14. The study of the PMS objects presented in this paper
refers to an area of $\sim 5\arcmin$ radius around the cluster's centre
, which includes the WFI, SOFI and Spitzer datasets. 

Most of the results and of the discussion in this work concern the
physical properties of PMS in stars in Tr\,14 as derived from the WFI
catalogue at visible wavelengths, but we will also make use of the SOFI
and Spitzer near-IR catalogues in Section\,7, where we discuss
the complete spectral energy distribution (SED) of these objects. In a
forthcoming paper (Beccari et al. 2014, in preparation), we will present
a complete study of all PMS stars over the entire field covered by the
observations,  including also those in Trumpler\,16 and Collinder\,232.

\subsection{WFI observations} 

We retrieved from the ESO archives a set of multi-band observations
made with the WFI on 2003, June 27 (proposal 069.C-0426(C);
principal investigator J. Alves). The WFI consists of a mosaic of eight
CCD chips with a global field of view (FoV) of 
$\sim33\arcmin\times34\arcmin$ and with a pixel size of $0\farcs238$.
The Carina Nebula was observed in the $U$, $B$, $V$, $R$, and $I$ bands
and through the  $H\alpha$ narrow  band filter under good seeing
conditions ($0\farcs8-1\farcs2$). The details of the observations are
listed in Table~\ref{tab_obs}. The observations were made in such a
way to place the luminous blue variable $\eta$\,Car inside chip \#56  of
the camera array and to include the star clusters Tr\,14, Trumpler\,16
and Collinder\,232 inside the field of view.

We corrected the raw WFI images for bias and flat field, and the
overscan region was trimmed using standard IRAF\footnote{IRAF is distributed
by the National Optical Astronomy Observatories, which is operated by
the  Association of Universities for Research in Astronomy, Inc., under
contract to the National Science Foundation} tools. The photometry
was carried out using the standard point spread function (PSF) fitting
independently on each image using the DAOPHOTII/ALLSTAR 
routines~\citep[][]{ste87}. A master list was created using the stars
detected in at least six of the $V$, $R$ and $I$-band images. This
strategy allowed us to remove from the final catalogue spurious
detections such as cosmic ray hits or star-like peaks on haloes and
spikes around saturated stars.  The master list was used as input for
ALLFRAME~\citep[][]{ste94}. The average of the single frame's magnitudes 
for each object in the master list was adopted as
stellar magnitude in the final catalogue, while we used the standard
deviation as associated photometric uncertainty. We looked for variable stars 
in the final photometric catalogue but,  within the photometric
uncertainties, no systematic variability was observed between these
relatively short exposures.

\begin{table}
\caption{Log of the WFI observations. Together with the band and filter name, 
the table lists the central wavelength (CW) and full width at half
maximum (FWHM) of the filters, the number of exposures and their total
duration.}              
\label{tab_obs}      
\centering                          
\begin{tabular}{ l c c c l}        
\hline\hline                 
Filter & CW & FWHM & N exp. & exp. time  \\    
  & [nm] & [nm] & & [s] \\
\hline                        
 U50\_ESO877 (U)            & 340.4 &  73.2 & 4 & ~200 \\      
  B123\_ESO878 (B)            & 451.1 & 133.5 & 4 & ~200 \\
 V89\_ESO843 (V)            & 539.6 &  89.4 & 4 & ~200 \\
 RC162\_ESO844 (R)          & 651.7 & 162.2 & 4 & ~200 \\
  I203\_ESO879 (I)           & 826.9 & 203.0 & 4 & ~120 \\
 H$\alpha$7\_ESO856  (H$\alpha$) & 658.8&   7.4 & 4 & 1200 \\
\hline                                   
\end{tabular}
\end{table}

The instrumental positions, in pixels, were transformed into J2000
celestial coordinates by means of an astrometric solution  (in the form
of a second degree polynomial) obtained with
CataXcorr\footnote{The software is available for download
at http://davide2.bo.astro.it/~paolo/Main/CataPack.html} from more 
than 30\,000 stars in common  between our
final WFI catalog and the 2MASS  catalogue of the same region. The
r.m.s. scatter of the solution is less than $0\farcs3$ in both right
ascension (RA) and declination (DEC).

The calibration of the instrumental magnitudes was done following
the procedure described in~\citet[][hereafter D09]{da09}. As shown in
that work, the transformations  between the WFI instrumental photometric
system and the standard Johnson--Cousins system are non-linear,
especially when a large color (i.e. temperature) range is considered. 
We thus decided to not apply any color correction to our
instrumental magnitudes and to keep them in the WFI photometric system,
as defined by the throughput of the adopted WFI filters. Given the
instrumental magnitude in a given band $m_{\rm WFI}^{\rm ins}$,
corrected for the air mass of the science observations, the absolute
photometric calibration in the WFI-VegaMag photometric system will be
derived as $m_{\rm WFI}^{\rm vega}=m_{\rm WFI}^{\rm ins}+Zp$, where $Zp$
is the photometric zero point. In practice, $Zp$ is the value one would
obtain for a Vega-like star, when all the colors are equal to zero (see 
D09 for details).

We derived the value of $Zp$ for each band by using a set of photometric
standards acquired during the night of the observations. Since the $Zp$
values that we calculated are identical, within the uncertainties, to
those derived in 2003 March by the WFI support team on a photometric
night  and published by the observatory
\footnote{http://www.eso.org/sci/facilities/lasilla/instruments/wfi/inst/zeropoints/},
we used the latter to calibrate our data. Since the $Zp$ value is not
available in that list for the H$\alpha$
band, we derived it directly with the procedure indicated above, 
obtaining $Zp(H\alpha)= 21.630 \pm 0.096$.

\subsection{SOFI observations}

We used a set of near-IR observations
made with the SOFI camera in "large field" mode on the night of
2 May 2008 (proposal 078.D-0371(A); principal investigator C. Weidner). In this
mode the camera has a pixel scale of $0\farcs288$,  allowing  us to
sample a region of $4\farcm9 \times 4\farcm9$ around the centre of the
cluster (see Figure\,\ref{fig_fov}). Exposures through the $J$ and $K$
filters were obtained with slightly different detector setups, namely 
DIT$=$4~s and NDIT$=$10 for the $J$ band and DIT$=$4~s and NDIT$=$20
in $K$. The same exposure was repeated 20 times in $J$
and 40 times in $K$ with a random dithering pattern for background
correction purpose.  During the observations the average seeing was
$\sim 1\arcsec$.

Images were pre-reduced using standard IRAF routines. A sky image,
obtained by  median combination of the dithered frames of each filter,
was subtracted from each frame. Flat fielding was then done using
the {\it SpecialDomeFlat}  template, which applies the appropriate
illumination corrections, as described in the SOFI User Manual~\citep[][]{li00}. 
Finally, for each filter all the dithered frames obtained  in a sequence were
aligned in coordinates and then averaged in a single image. Standard
crowded field photometry, including PSF modeling, was then carried out
on each of the averaged frames using DAOPHOTII/ALLSTAR.

We used hundreds of stars in common with the 2MASS catalogue in order to
obtain an absolute astrometric solution for the stars in the SOFI
catalogue. The same stars where also used as secondary photometric
standards in order to calibrate the SOFI $J$ and $K$ magnitudes with the
2MASS photometry.

\subsection{Spitzer observations}

Finally, we retrieved from the Spitzer Enhanced Imaging Products
archive\footnote{http://irsa.ipac.caltech.edu/data/SPITZER/Enhanced/Imaging/index.html}
two near-IR images (at $3.6\,\mu$m and $4.5\,\mu$m) centered on Tr\,14,
obtained with the IRAC camera (proposal 30734; principal investigator D. Figer). 
The two mosaic images are the combination of a number of
sub-images and have a pixel scale of $0\farcs6$. Standard PSF photometry
was performed by modeling the PSF on each image, producing 
a catalogue listing the relative positions and magnitudes of stars in
common  between the two bands.

The calibration of the Spitzer observation was done following the
the IRAC instrument 
handbook\footnote{http://irsa.ipac.caltech.edu/data/SPITZER/docs/i}
\citep[see also][]{ho08}.
We performed aperture photometry with the IRAF/DAOPHOT task on more than 40
well sampled and isolated stars in each frame using an aperture of 4
pixel while the background was locally sampled in a annulus between 4
and 12 pixel, following the IRAC instrument handbook.  We used
these stars as secondary photometric standards to calibrate the final
photometric catalogue. Moreover, several hundreds stars in
common with the SOFI catalogue were used to extend the astrometric
solution to the Spitzer catalogue.

\section{Differential reddening}
\label{sec_redd}

Since the stars in our sample are still embedded in the molecular cloud,
a large amount of differential reddening is expected to affect the
photometry. Recently H12 carried out a detailed
study of the differential reddening in the same region, finding that
inside the cluster the extinction law is characterized by  $R_{V}^{\rm
cl}=4.4\pm0.2$. They also determined the foreground reddening to be
$E(B-V)_{\rm fg}=0.36\pm0.04$ for the stars in the $\eta$ Carinae nebula,
based on the properties of 141 early-type stars with high  
membership probability from proper motion studies. They derived a distance modulus of
$12.3\pm0.2$, corresponding to a distance of $2.9 \pm 0.3$\,kpc for
Tr\,14 and Tr\,16 and concluded that the two clusters are at the same
distance. Finally, they derived a reddening map (see their figure 6)
using the early-type stars in the region, and used it to correct the
magnitudes of the entire population, including those of PMS stars. 

Taking advantage of the large number of filters available in our
photometry, we decided to independently derive a reddening map using
the method described in~\citet[]{mi81} and \citet[hereafter
R02]{rom02}. We adopted the reddening free color $Q_{UBI}$ defined as
follow:

$$
Q_{UBI}=(U-B)-\frac{E(U-B)}{E(B-I)}(B-I)
$$

\begin{figure} 
\centering 
\includegraphics[width=0.5\textwidth]{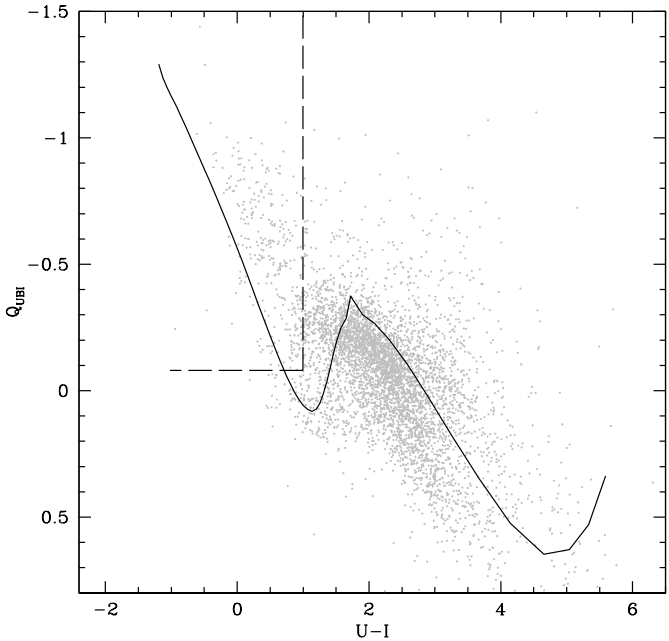}
\caption{The solid line shows the reddening free color $Q_{UBI}$ as a 
function of $U-I$ as derived from the model atmospheres
of~\citet[][]{be98} and reddened by the foreground extinction (see details in the text), 
whereas the grey dots correspond to the stars in our
catalogue. Only stars with $U-I<1$ and $Q_{UBI}<-0.1$ (area inside the
dashed lines) were used to generate our reddening map.}
\label{fig_red}
\end{figure}

The reddening-free color $Q_{UBI}$ computed for the observed stars (grey points)
and for the atmospheric model of~\citet[][; solid line]{be98} is shown in Figure~\ref{fig_red} as
a function of $(U-I)$. The theoretical model shown in the figure is already reddened by the foreground extinction
as found by H12.

The effect of extinction is to move the stars horizontally from the 
zero-reddening locus, i.e. the location of the model in the colour--colour 
diagram. We can simply derive information on the interstellar reddening by comparing the 
colors of the observed stars with those of the models, in the assumption that the displacement 
of the stars from the W-shaped model is only induced by the differential extinction. 

As shown in Figure~\ref{fig_red}, only the stars with $U-I<1$ and
$Q_{UBI}<-0.1$ do show an unique solution. Moreover, it's important to stress that the stars 
with $Q_{UBI}<-0.5$ and $U-I>1$ cannot be safely used to study the reddening map because this 
region of the diagram is mostly populated by PMS stars, whose colors are clearly 
affected by U-band excess. These limits define a ``selection box'' shown with the 
dashed lines in the figure. 
In order to build a reddening map for the entire region covered by our WFI observations
we first derived the value of visual extinction $A_V=R_V\times E(B-V)$ for the stars in the selection box.
While, as previously explained, $E(B-V)$ can be derived by projecting horizontally the stars to the 
model in the $Q_{UBI}$ vs $U-I$ plane of Figure~\ref{fig_red}, we need to assume an extinction
law $R_V$ in order to derive $A_V$.  Following the recent results from H12 we estimated
the total $A_V$ of the selected stars as $A_V = 3.1\times E(B-V)_{fg}+R_{V, cl}\times (E(B-V)-E(B-V)_{fg})$, 
where the only parameter missing is $E(B-V)$ that we estimate using the $Q_{UBI}$ parameter.

\begin{figure} \centering \includegraphics[width=0.5\textwidth]{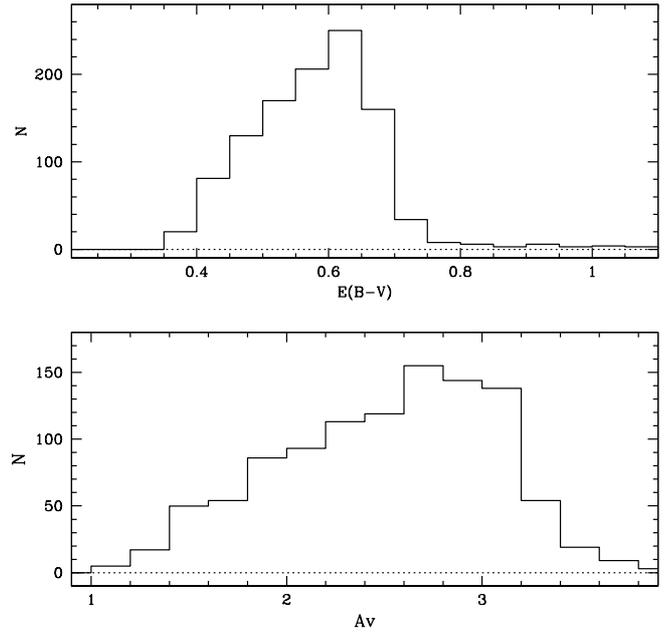}
\caption{E(B-V) (upper panel) and Av (lower panel) distribution as derived from the reddening free colors 
in the region sampled by the WFI data.}
\label{fig_av}
\end{figure}

In Figure~\ref{fig_av} we show the histograms of the values of the total
extinction $A_V$ (lower panel) and selective extinction $E(B-V)$ (upper
panel) as calculated using more than 1000 stars located within the 
selection box of Figure\,2. The median of the $E(B-V)$ distribution is $0.59 \pm 0.06$ with a
minimum value of 0.075 and a maximum value of 1.425 in agreement with
the findings of H12 and of~\citet[][]{ca04}. In order to produce an
extinction map, the entire FoV was divided into a grid of $3\arcmin
\times 3\arcmin$ cells, for a total of 64 cells. The extinction in each of them was obtained as
the average value of the extinction of the individual blue stars in that
cell and this value was assumed as representative for all stars
falling inside that cell. The typical standard deviation of the mean of $E(B-V)$ of the individual blue stars
in the cells in the area of 5\arcmin around the center of trumpler 14 (the region studied in detail in this paper) 
is $\sim0.08$, which translates into an uncertainty of $\sim0.35$ on $A_V$ since $R_V=4.4$.
We used this map to correct for extinction 
all the available magnitudes, from the optical through to the IR,
following the extinction law of H12. As we will show later and as
already pointed out by H12, the stars in the foreground will appear too
blue when corrected with this map and will need to be treated
separately.

\section{The color--magnitude diagram}
\label{sec_cmd}

\begin{figure} 
\centering 
\includegraphics[width=0.5\textwidth]{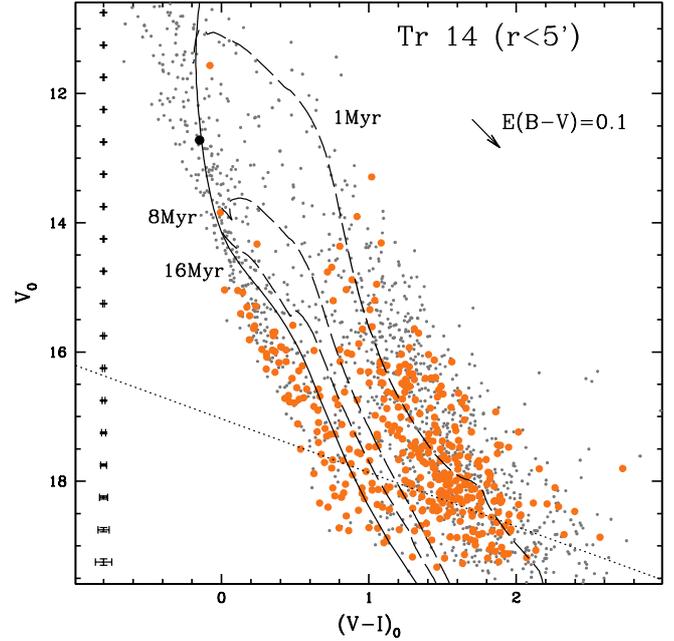}
\caption{CMD of all stars within $5\arcmin$ of the cluster centre, after 
correction for extinction.  Stars highlighted with a thick dot (orange in the electronic
version) are bona fide-PMS. The selection of this stars and 
the physical parameters are discussed later in the paper. Three PMS
isochrone~\citep[][]{de08} for ages of 1, 8 and 16 Myr are shown as reference. The ZAMS 
from~\citet[][]{ma08} for solar metallicity and the assumed distance modulus $(m-M)_0=12.3$ is shown by 
the solid line. The typical uncertainties on magnitudes and colors are 
indicated by the crosses. The dotted line shows the 50\,\% completeness
limit of the photometry. A reddening vector proportional to $E(B-V)=0.1$ 
(i.e. the   typical uncertainty) is also shown.}
\label{fig_cmd}
\end{figure}

The optical CMD of the inner 5\arcmin of Tr\,14 is shown in 
Figure~\ref{fig_cmd}. Stars highlighted with a thick dot (orange in the electronic
version) are bona fide-PMS. The
selection criteria and physical parameters for these class of object
will be discussed in detail in this section and in Section~\ref{sec_pms}, respectively.

All magnitudes are corrected for extinction on the
basis of our extinction map. Typical photometric uncertainties (on
magnitudes and colors) are shown by the crosses, for various
magnitudes. In the figure we also show the position of the zero age 
main sequence (ZAMS), taken from the models of~\citet[][]{ma08} for
solar metallicity (solid line). We have adopted as
distance modulus the value given by H12, namely $(m-M)_0=12.3$. The
position of the ZAMS obtained in this way helps us to identify a
population  of candidate MS stars that extend from the saturation limit
at $V_0 \simeq 11$ down to $V_0 \simeq 19$ . 

In order to measure the completeness of our photometry, we made
extensive use of artificial stars experiments, following  the recipe
described in~\citet[][]{be02}. More than 300\,000 artificial stars were
uniformly distributed on the WFI chips in groups
of 4\,000 stars at a time in order to avoid changing the crowding
conditions. We produced a catalogue of simulated stars with a $V$
magnitude, $V-I$ and $V-H\alpha$ colors randomly extracted from 
a luminosity function and colour distributions modeled to reproduce 
the observed one.

The whole data reduction  process was repeated as for the
actual observations, and the fraction of recovered stars was estimated
at each magnitude level. The limit of 50\% of photometric completeness
within $5\arcmin$ of the centre of Tr\,14 is shown by the
dotted line in Figure~\ref{fig_cmd}. From a comparison
with the theoretical model we estimate that we sample the PMS stats down to
0.7 M$_{\odot}$ and 1 M$_{\odot}$ with a
50\% photometric completeness, with the lower masses sampled at the younger ages. 

A discontinuity in the stellar color distribution in this CMD separates
a population of objects along the ZAMS from one clearly grouped  at
redder colors and consistent with the young population of PMS stars
already detected in Tr\,14 (H12; Ascenso et al. 2007). This confirms that
Tr\,14 is an active star-forming region. In order to learn more about
the properties of this recent star formation episode, we followed the
method developed by DM10 to identify {\it bona-fide} PMS stars. 
A population of candidate PMS stars is well visible at $V-I \ga 1$,
well  separated from the population of MS stars. These PMS stars have
been recently studied by H12, who selected them based on their position
in the optical CMD. In this work, we apply the method
developed by DM10, to look for the signature of the active  mass accretion
process that is expected to characterize the PMS phase and that is 
responsible for the strong H$\alpha$ excess emission normally observed
in objects of this type~\citep[e.g.][]{ca00}. 

\begin{figure} 
\centering 
\includegraphics[width=0.5\textwidth]{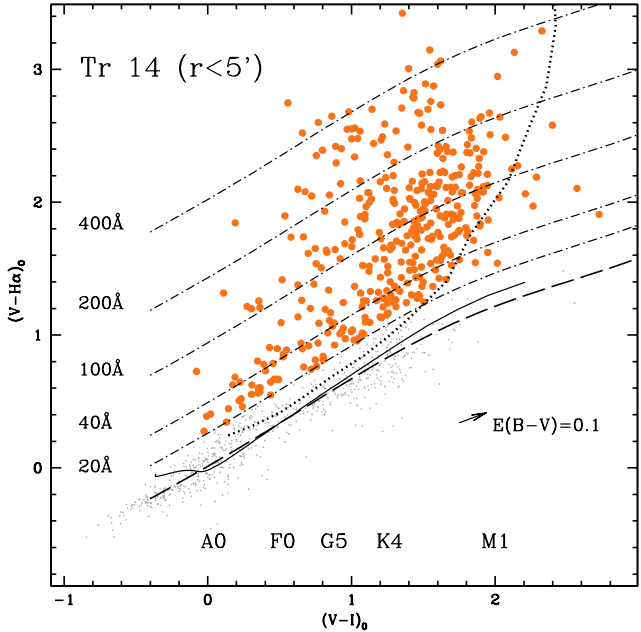}
\caption{Color-color diagram of all stars after correction for reddening. Stars in light 
grey correspond to objects sampled in the first $5\arcmin$ from the 
cluster centre. The solid line represents the median (V-H$\alpha$)$_0$ colour of stars 
with an error on  (V-H$\alpha$)$_0<0.05$, and is defined as the locus of stars without 
H$\alpha$ excess emission and hence the location of stars with EW(H$\alpha$)=0.
Objects indicated as thick dot (orange in the electronic version) correspond to 
stars with H$\alpha$ excess emission. The solid line shows the location 
on the CMD of atmospheric models of~\citet[][]{be98}. The model is in excellent
agreement with the observations. The dotted line marks the 50\% completeness limit of the photometry.
Spectral types are also indicated. A reddening vector proportional to $E(B-V)=0.1$ 
(i.e. the   typical uncertainty) is also shown.}
\label{fig_viha}
\end{figure}

In Figure~\ref{fig_viha} we show the distribution of the stars in the
$V-I$ {\em vs.} $V-H\alpha$ color--color diagram, corrected for
extinction. We use the median $(V-H\alpha)^{ref}$ de-reddened color of
stars with small ($<0.05$ mag) combined photometric uncertainty in the
$V$, $I$ and $H\alpha$ bands (grey dots in Figure~\ref{fig_viha}),  as a
function of $V-I$, to define the reference template with respect to
which the excess H$\alpha$ emission is identified (dashed line). 

We selected a first sample of stars with excess H$\alpha$ emission by
considering  all those with a
$\Delta(H\alpha)=(V-H\alpha)^{star}-(V-H\alpha)^{ref}$ at least 4 times 
larger than the photometric uncertainty on the $(V-H\alpha)^{star}$
color. Then we calculated the equivalent width of the H$\alpha$ 
emission line, $EW(H\alpha)$,  from the measured color excess using
Equation\,4 of D10. We finally considered as {\it bona-fide} PMS stars
those objects with EW(H$\alpha$) $>20\AA$ (see D10 and references
therein). This allows us to safely remove from our sample
possible  contaminants, such as older stars with chromospheric activity
and Ae/Be stars~\citep[see][]{wb03, be14}.

We visually inspected the position of all the candidate H$\alpha$
emitters on the $V$, $I$ and $H\alpha$ images and excluded from the
final list all those stars falling on a filamentary structure of the
cloud in which the cluster is embedded. Even if these stars could in
principle be genuine H$\alpha$ excess sources (and hence considered PMS
stars), the fact that the scale of the background spatial variation is equal or smaller
than the full width at half maximum of the PSF makes the background
determination too prone to errors. As a consequence, the magnitudes of
these stars is too uncertain and we do not consider them as {\it
bona-fide} PMS objects. After this conservative  selection, we count a total of 
389 PMS stars in our sample, with a median $EW(H\alpha)=90$\,\AA.
Hereafter we will refer to these objects as {\it bona-fide} PMS stars.

The positions of these stars in the optical CMD is shown in
Figure~\ref{fig_cmd} together with a ZAMS from~\citet[][solid line]{ma08}
and three PMS isochrone for ages of 1, 8 and 16 Myr from~~\citet[][dashed lines]{de08}
for solar metallicity.

\begin{figure} 
\centering 
\includegraphics[width=0.5\textwidth]{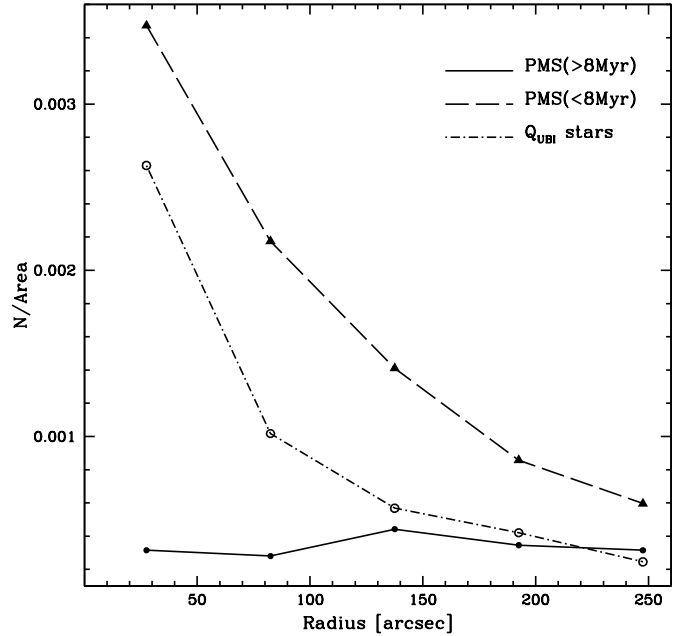}
\caption{Radial density distribution of old (age$>8$Myr) and young (age$<8$Myr)
bona-fide PMS together with the stars used in  Section\ref{sec_redd} to calculate
the reddening map of the region.}
\label{fig_rad_iniz}
\end{figure}

\subsection{The reddening correction of the old PMS stellar population}
\label{sec_old_pms}

Using the same approach as described by \citet[][]{de11b}, we  
perform a preliminary study of the radial distribution
of the candidate PMS stars at different ages. In particular, using the
8\,Myr isochrone
as references, we divided the population of bona-fide PMS in an older
($>8$\,Myr)
and younger ($<8$\,Myr) subgroups. In Figure~\ref{fig_rad_iniz} we show the
radial density distribution of the two subgroups of PMS objects, with
respect to the centre of Tr\,14. The solid and dashed lines correspond,
respectively, to the older and younger populations, while the
distribution of upper MS stars used in Section~\ref{sec_redd} to derive
the reddening map of the region is shown by the dot-dashed line. 
The figure clearly shows that objects with H$\alpha$ 
excess older than $\sim 8$ Myr have a different radial distribution from 
that of the cluster's stars used to calculate the reddening map. 

The implication of the differences in the radial distribution of the PMS populations
will be discussed in more details in Section~\ref{sec_pms}.
For the moment, Figure~\ref{fig_rad_iniz} implies that we can safely use the reddening map
to correct the magnitudes of the younger bona-fide PMS objects for differential reddening, by
assuming that the placement of these PMS objects with respect to the 
cluster's absorbing material is similar to that of the reference stars in its vicinities.
This assumption is clearly not valid for the older H$\alpha$ excess emitters.
We also note that the older PMS stars are not in regions of particularly high reddening, 
confirming that they should have lower extinction than younger PMS stars. At the same time, as
explained in Section~\ref{sec_redd}, the typical uncertainty on $E(B-V)$ is $\sim0.1$ 
It is then clear that the reddening uncertainty has a marginal effect on the distribution of the stars in the diagrams.

In order to obtain a more realistic estimate of the reddening of the older population of PMS, we can use the stars with H$\alpha$
excess that appear bluer than the ZAMS (e.g. Figure~\ref{fig_cmd}). By definition the ZAMS represents the theoretical
location of the stars at the beginning of the hydrogen burning phase in the core, i.e. the MS stage, or
conversely, the end of the PMS phase. Observationally, the ZAMS represents the bluest side
of the colour distribution of the PMS population in a CMD. We then decided to correct the
magnitudes of the older population of PMS stars using a constant value 
of $A_V$=1.7, which ensures that the ZAMS constraints the blue envelope to
at least 80\% of the old PMS objects at $V_0 \simeq 16 - 18$ 
in Figure~\ref{fig_cmd}, i.e. to the stars in this subgroup with the
smallest uncertainty on magnitude and colour. For the faintest stars, the increase of the photometric uncertainties 
can easily be at the origin of some residual colour scatter around the ZAMS.
While the choice of 80\% is a somewhat arbitrary, it is important to stress here that
changing this value to 50\% or 100\% would not change any of the results of the paper.

\begin{figure} 
\centering 
\includegraphics[width=0.5\textwidth]{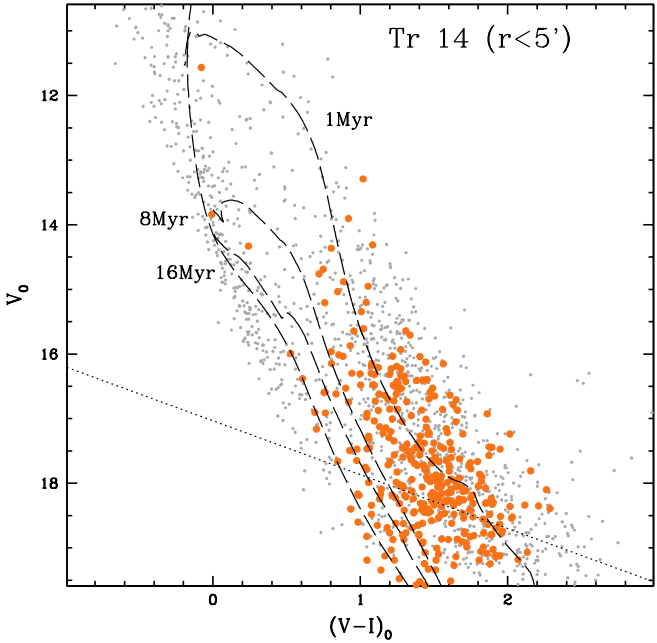}
\caption{CMD of all stars within $5\arcmin$ of the cluster centre, after 
correction for extinction. The symbol types are the same as used in 
Figure~\ref{fig_cmd}. A reddening correction with a constant $A_V$=1.7 has 
been used for the PMS stars older than 8\,Myr in Figure~\ref{fig_cmd}. The 
dotted line marks the 50\,\% completeness limit of the photometry.}
\label{fig_cmd2}
\end{figure}

Finally, we revised the selection of the H$\alpha$ emitters (bona-fide PMS stars), In fact, even if 
the reddening vector runs almost parallel to the median photospheric colors of normal stars
in the $V-H\alpha$ vs $V-I$ colour--colour diagram~\citep[see][]{de11b},
we calculated the new values of $EW(H\alpha)$ after the application of the new reddening 
correction to the $V$, $I$ and $H\alpha$ magnitudes of the old PMS objects. We find that 33 stars 
now show  $EW(H\alpha)<20\,\AA$ and, following our stringent selection criteria already
reported earlier in this section, these stars can not be classified as H$\alpha$ excess
emitters and will no further be considered in this work. 
The population of bona-fide PMS stars in Tr\,14 finally includes 356 objects.
 
\section{Physical parameters of the PMS stars}
\label{sec_pms}

\begin{figure} 
\centering 
\includegraphics[width=0.5\textwidth]{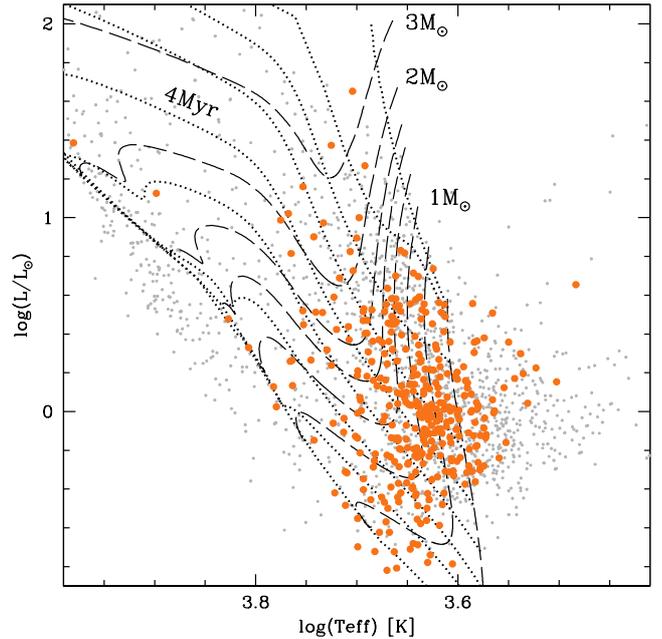}
\caption{All objects located within $5\arcmin$ of the cluster centre are 
shown in grey in the H--R diagram. Bona-fide PMS stars are shown as
thick dots (orange in the electronic version). Dashed lines show the evolutionary tracks for solar 
metallicity and masses of $0.6$, $0.8$, 1, $1.2$, $1.4$, $1.6$, 2, and 3 
M$_{\odot}$. The corresponding isochrones are shown as dotted lines, for 
ages of $0.25$, $0.5$, 1, 2, 4, 8, 16, and 32 Myr, from right to left. 
The constant logarithmic age step has been selected in such a way that 
the typical distance between isochrones is larger than the photometric 
uncertainties. }
\label{fig_hr}
\end{figure}

The luminosities ($L$) and effective temperatures ($T_{\rm eff}$) of all
stars within $5\arcmin$ of the cluster centre are shown in the
H--R diagram (Figure~\ref{fig_hr}). PMS stars are shown as thick  
dots (orange in the electronic version), whereas all other objects are in grey. 

Effective temperatures $T_{\rm eff}$ were determined by fitting the
de-reddened $(V-I)_0$ colors of the stars with the ones  computed for the
same WFI bands using stellar atmosphere models. We adopted the model
atmospheres of~\citet[][]{be98}, which provide an excellent fit to our
data (solid line in Figure~\ref{fig_viha}).  The de-reddened magnitudes
were then compared with the absolute magnitudes of the models in order
to derive the stellar radius. The bolometric luminosity was finally
derived, adopting the distance modulus $(m-M)_0=12.3 \pm 0.2$
from H12 and by comparing the stellar radius and $T_{\rm
eff}$ with those of the Sun \citep[see][for further details]{rom02}.

We also show in Figure~\ref{fig_hr} the evolutionary tracks for masses
of $0.6$, $0.8$, 1, $1.2$, $1.4$, $1.6$, 2, and 3 M$_{\odot}$ (dashed lines) and isochrones for ages of $0.25$, $0.5$, 1, 2, 4, 8, 16,
and 32\,Myr  (dotted lines), from the  PMS evolutionary models of the
Pisa group~\citep[][]{de08,to12}. The models were calculated for Z =
0.014, Y = 0.2533 and a mixing-length parameter $\alpha=1.68$,
applicable to Tr\,14. In Figure~\ref{fig_hr} we  explicitly label the tracks
for 1, 2 and 3 M$_{\odot}$ and the location of the 4\,Myr
isochrone, for reference.  

\begin{figure}  
\centering 
\includegraphics[width=0.5\textwidth]{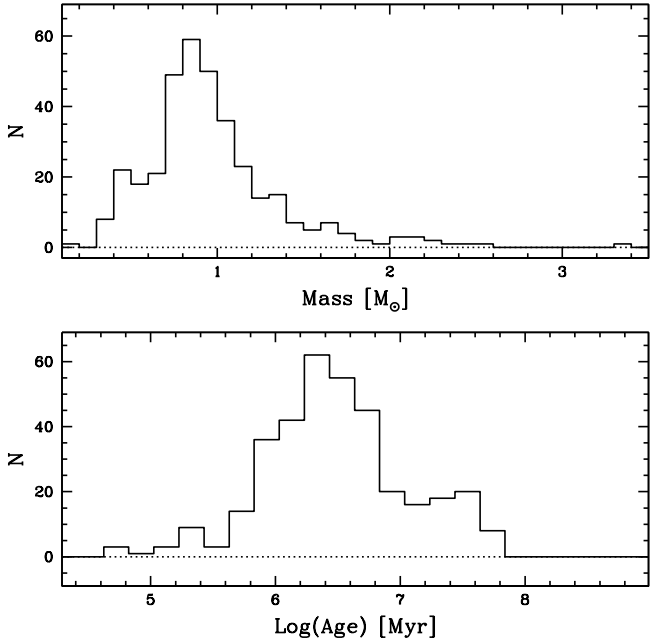} 
\caption{Mass and age distribution of the bona-fide PMS stars (upper 
and lower panel, respectively).} 
\label{fig_histo} 
\end{figure}

In order to study in more deal the physical properties of the 356 PMS in Tr14,
we used a finer grid of models than the one shown in Figure~\ref{fig_hr} to derive a reliable
measure of the mass and age~\citep[see][for details on the method used, with an approach similar to
the one more recently presented  by Da Rio et al. 2012]{rom98}. Mass and age 
distributions for the PMS objects are shown in the
upper and lower panels of Figure~\ref{fig_histo}, respectively.

The mass distribution peaks around 1 M$_{\odot}$. As expected the number
of stars with masses above $\sim 1$\,M$_{\odot}$ decreases, being the 
timescales of PMS evolution faster for more massive objects. On the 
other hand, the decrease of the number of stars for masses below
1\,M$_{\odot}$ is due to the drop of the photometric completeness see
the dotted line in Figure\,\ref{fig_cmd2}). 

Concerning the ages, the median of the distribution is $\sim 2.8$\,Myr,
with an upper quartile of $\sim 6$\,Myr and a lower quartile of $\sim
1.2$\,Myr. As already noticed by~\citet[][]{as07}, it appears that star
formation has been ongoing in Tr\,14 for at least 5--6\,Myr. The available
observations do not allow us to conclude whether the star formation in
the last 6 Myr was continuous or whether this young population of PMS
objects is indeed the outcome of a few major bursts, as postulated
by~\citet[][]{as07}.

On the other hand, it is clear that multiple bursts have occurred in
this region. The distribution of PMS ages confirms that $\sim 20\,\%$ of
the objects have ages older than 8\,Myr, with a median age of $\sim20$Myr. 
This implies that in this region star formation started more than 8 Myr ago, at 
least as far back as $\sim 20$\,Myr ago, followed by a major episode in the last
2--6\,Myr. A similar behavior has already been observed in Galactic star forming 
regions (NGC\,3603, Beccari et al. 2010; NGC\,6823, Riaz et al. 2012; 
NGC\,6611,  De Marchi et al. 2013b) and in the Magellanic 
Clouds~\citep[][]{pa00, de10,de11a,de11b,spe12,de13a}.

Our age resolution does not allow us to say whether there
were multiple episodes in Tr14 before the most recent one, nor whether they were
as prominent as the one responsible for the 2--4\,Myr old population.
However, if the fraction of PMS stars with H$\alpha$ excess has an
exponential decay with time (Fedele et al. 2010), episodes occurring
more than 10\,Myr ago might have contributed a large portion of the
stars in this region. 

It is important to recall that there are a number of uncertainties when
searching  for age spreads in starburst clusters~\citep[see e.g. the reviews
by][and Soderblom et al., 2013]{pr12}. From a theoretical point of view, while the absolute ages
of PMS stars are only as good as the isochrones allow, relative ages are
always  better defined than a factor of two for ages up to $\sim
30$\,Myr, as already shown by~\citet[][]{rom02} and~\citet[][]{da10}. 
From an observational point of view, the colors and magnitudes of the
stars are affected by uncertainties induced by inaccurate evaluation of
differential extinction, stellar variability~\citep[][]{he02} and
complex physical processes like episodic accretion~\citep[][]{ba09}. We stress
once again here that the typical uncertainty on $E(B-V)$ in the 
studied region is of the order of 0.08 mag, i.e. $\sim0.35$ on $A_V$. 
Hence, while an inaccurate estimate of $A_V$ would have an impact on the age 
determination of individual stars, we can exclude that it would have a significant 
impact in the age spread detected in the region.
Unresolved binaries can also affect the apparent distribution of stars in the
CMD and lead to an incorrect estimation of the age of such photometrically
unresolved system. We stress anyway that the contamination for stars 
appearing older than 10 Myr should not be more than about 5\% 
~\citep[for 100\% unresolved binaries;][]{we09}. 
However, none of these effects  can explain the different radial
distributions of younger and older PMS stars as shown in a number of
star-burst clusters~\citep[e.g.][]{be10,de11a,de11b,de13a,de13b}.

\begin{figure} 
\centering 
\includegraphics[width=0.5\textwidth]{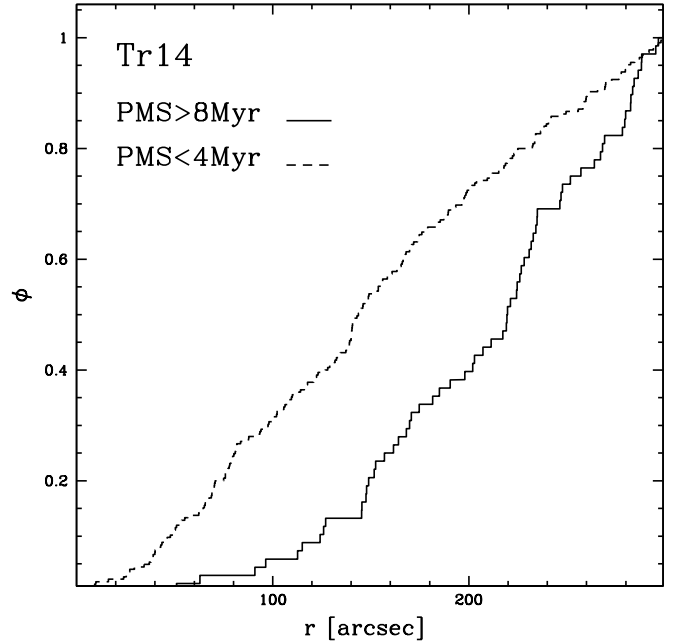}
\caption{Cumulative radial distribution of the younger ($< 4$\,Myr;
dashed line) and older ($> 8$\,Myr, solid line) PMS objects in Tr\,14.}
\label{fig_ks}
\end{figure}

Figure\,\ref{fig_rad_iniz} already showed that this is also the case for
Tr\,14. To characterize this result in an even more solid statistical
sense, in Figure~\ref{fig_ks} we compare the radial distribution of the older PMS stars ($>
8$\,Myr, solid line) with that of the PMS stars younger than 4\,Myr
(dashed line), with respect to the cluster centre. The latter age was
selected to ensure that most of these stars have the same age as the
massive objects that define the cluster, i.e. the O and B stars. 

The figure clearly shows that the younger PMS stars are more centrally
concentrated than the older generation, in agreement with the results
found e.g. in NGC\,3603 and NGC\,6611 in the Galaxy \citep[][]{be10,
de13b} and in 30\,Doradus, NGC\,346 and NGC\,602 in the Magellanic
Clouds~\citep[see][respectively]{de11a,de11b, de13a}. We used a
Kolmogorov--Smirnov (K--S) test to check the statistical significance of
the differences in the observed distributions. The test yields a 
confidence level exceeding $3\,\sigma$ that the two groups of stars have
a different radial distributions, supporting the hypothesis that the two
classes of PMS objects belong indeed to different generations of stars. 

\section{Mass accretion rates}

The $\Delta(H\alpha)$ parameter defined in section~\ref{sec_cmd} allows
us to easily derive the $H\alpha$ emission line luminosity $L(H\alpha)$
for our bona-fide PMS stars. As explained in~DM10, $L(H\alpha)$  is
obtained from $\Delta(H\alpha)$, from the photometric zero point and
absolute sensitivity of the instrumental  set-up and from the distance
to the sources. We have assumed the $Zp(H\alpha)= 21.63$ and a distance
modulus $(m-M)_0$=12.3, as mentioned before.  The median $L(H\alpha)$ in
our sample is $\sim 9 \times 10^{-3}$\,L$\odot$ with the 17 and  83
percentile levels in the distribution at $0.006$\,L$\odot$ and
$0.018$\,L$\odot$, respectively.

Following DM10, we derived the accretion luminosity $L_{acc}$ from the
measured value of  L(H$\alpha$). In that work it is shown that the ratio
$L_{acc}/L(H\alpha)$ can be assumed to be constant and, from an 
elementary fit to the data in the compilation of~\citet[][]{da08} of PMS
stars in the Taurus--Auriga association, it is found that $\log
L_{acc}=(1.72 \pm 0.25)+\log L(H\alpha)$. The median  value of $L_{acc}$
that we find for the PMS stars in Tr\,14 is $\sim 0.3\, L_\odot$. 

The mass accretion rate $\dot{M}_{acc}$ is in turn related to $L_{acc}$
via the free-fall equation, linking  the luminosity released by the
impact of the accretion flow with the rate of mass accretion, according
to the relationship:

\begin{centering}
\begin{equation}
L_{acc}=\frac{G \, M_* \, \dot{M}_{acc}}{R_*} 
\left(1-\frac{R_*}{R_{\rm in}}\right)
\label{eq_lacc}
\end{equation}
\end{centering}

\noindent where $G$ is the gravitational constant, $M_*$ and $R_*$ the
stellar  mass and photospheric radius derived in Section\,5, while
$R_{\rm in}$  is the inner radius of the accretion disc. The latter is
rather uncertain  and depends on the exact coupling between the 
accretion disc and the  magnetic field of the star. Following~\citet[][]{gu98}
we assume $R_{in}=5R_*$ for all PMS objects.

\begin{figure} 
\centering 
\includegraphics[width=0.5\textwidth]{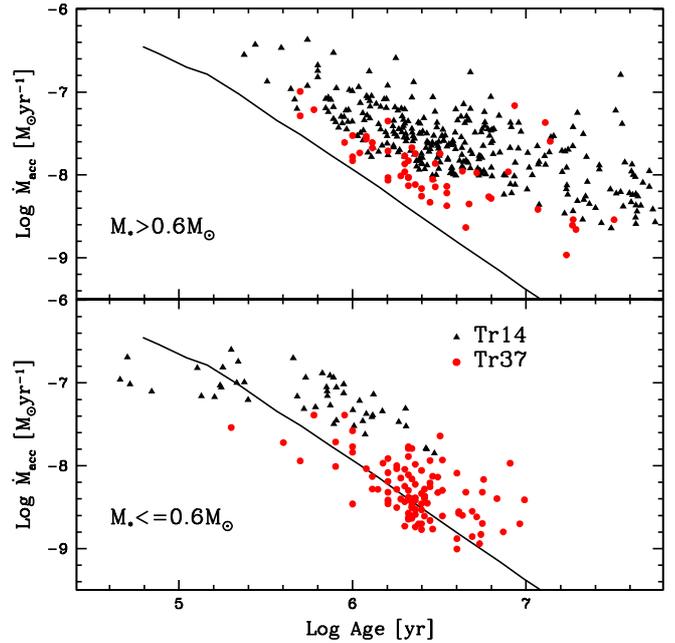}
\caption{Derived mass accretion rates as a function of the age of the 
PMS stars. Stars with masses larger and smaller than $0.6$\,M$_\odot$
are shown separately. The thick black dots are for the stars in Tr\,14,
whereas the smaller grey dots correspond to objects in Tr\,37. The solid
line shows the temporal evolution of $\dot{M}_{acc}$ for viscous discs
as predicted  by~\citet[][]{ha98}. 
}
\label{fig_disc}
\end{figure}

The values of $\dot M_{\rm acc}$ derived in this way for the PMS stars in Tr\,14 are shown as black triangles in 
Figure~\ref{fig_disc}, as a function of the stellar age. The solid line shows the temporal decline of $\dot M_{\rm acc}$ for the 
evolution of viscous discs, as predicted by~\citet[][]{ha98} for stars of mass $\sim 0.4$\,M$_\odot$. It is clear that the 
distribution of $\dot M_{\rm acc}$  for the PMS objects in our sample is much higher compared to the theoretical predictions. 
~\citet[][]{da14} have shown that correlated uncertainties between $\dot M_{\rm acc}$ and the estimated 
parameters of young stars (such as luminosity, effective temperature, mass, age) can introduce biases on the 
apparent decay of $\dot M_{\rm acc}$ as a function of age. More generally, diagrams like those shown in Figure~\ref{fig_disc} 
can be used to study the temporal evolution of $\dot M_{\rm acc}$ only if all stars, also the older ones, were formed 
under similar conditions. To avoid these possible sources of uncertainty, we concentrate on stars of similar age, $\sim 2$ \,Myr, 
which are numerous in Tr\,14, as well as in the other cluster to which we will compare it later (Tr\,37). The mean value 
of $\dot M_{\rm acc}$ at an age of $\sim 2$\,Myr is $\sim 1.6 \times 10^{-8} M_{\odot}yr^{-1}$, while the theoretical model predicts 
a value of $\sim 5 \times 10^{-9} M_{\odot}yr^{-1}$.

In order to better understand the origin of this apparent discrepancy we
used the photometric catalogue of T-Taury stars belonging to the H\,II 
region IC\,1396 in Cepheus OB2, including Trumpler\,37 (Tr\,37),
published by~\citet[][hereafter B11]{ba11}.  In that paper the authors
use observations acquired with the Wide Field Camera at the Isaac Newton
Telescope (INT) in the  $r',g'$ and $H\alpha$ bands, as part of the INT
Photometric H-Alpha Survey (IPHAS; Drew et al. 2005). B11 used an
observational strategy very similar to the one used in this
paper to identify PMS stars from their H$\alpha$ excess emission and to 
study their physical parameters and $\dot{M}_{acc}$. We retrieved their 
catalogue of $r',i'$ and $H\alpha$ magnitudes, which also contains the
masses and radii derived for these objects. We followed an approach
described above to determine EW$(H\alpha)$, $L(H\alpha)$, $L_{acc}$ and
$\dot{M}_{acc}$ for the stars in the B11 catalogue. The values of 
$\dot{M}_{acc}$ found in this way are shown as red dots in
Figure~\ref{fig_disc}. The values are in full agreement with those found
by B11 (see their Figure\,16), confirming the validity of our approach.

\begin{figure} 
\centering 
\includegraphics[width=0.5\textwidth]{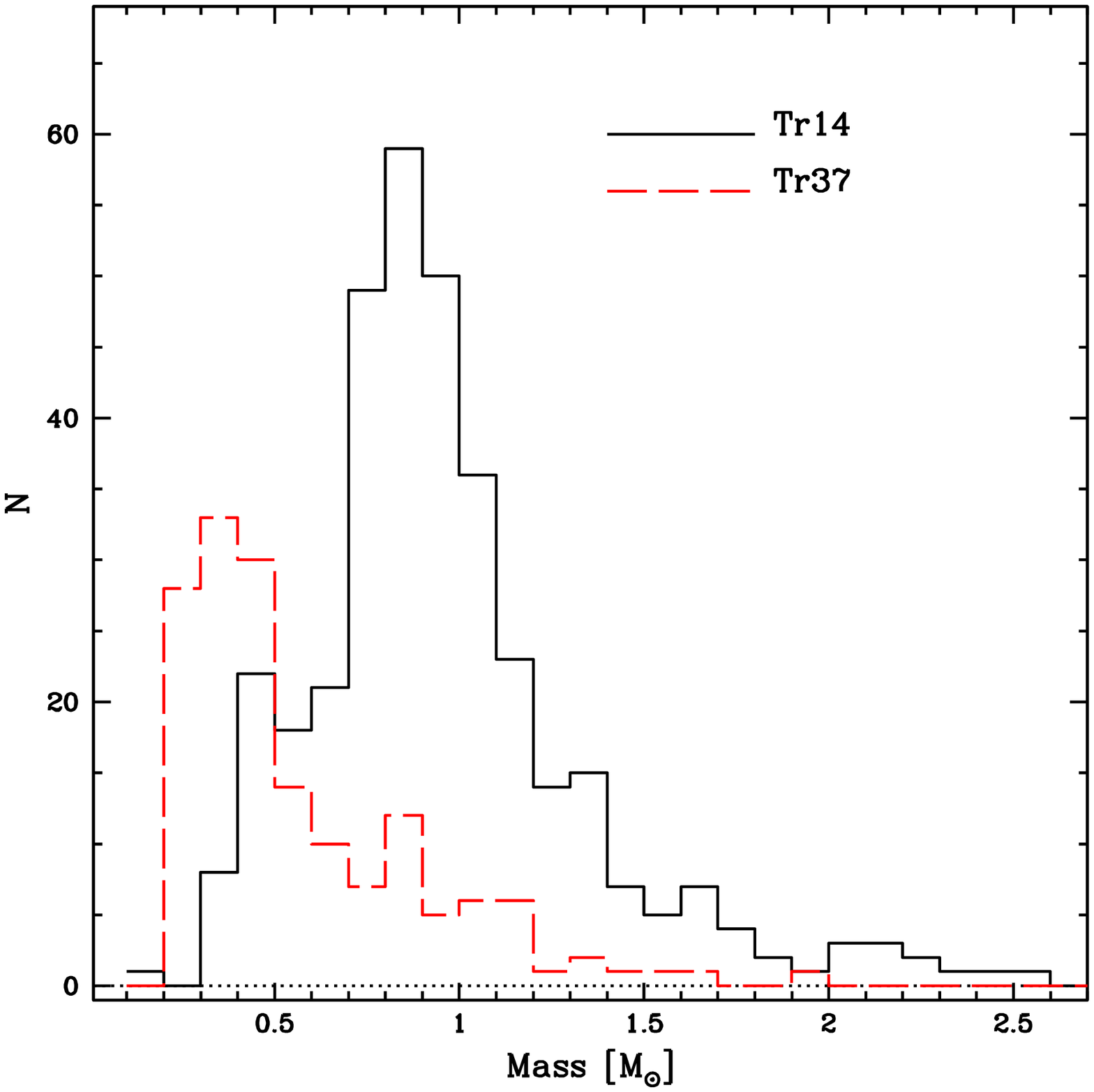}
\caption{Histograms showing the mass distributions of the PMS stars in 
Tr\,14 (solid line) and Tr\,37 from the catalogue of B11 (dashed line).}
\label{fig_comp}
\end{figure}

From the results shown in Figure~\ref{fig_disc} it is clear that the
$\dot{M}_{acc}$ of the stars in Tr\,14 are on average higher than those
measured in Tr\,37, which are more in agreement with the prediction form
the viscous model. In fact, this is only effectively the case for the
lower mass stars, shown in the lower panel of Figure~\ref{fig_disc},
while the objects more massive than $0.6$\,M$_\odot$ even in Tr\,37 have
$\dot{M}_{acc}$ values systematically above those predicted by the
models of \citet[][]{ha98}. To better characterise the differences
between the stars that we have sampled in the two regions, we compare
their mass distributions in Figure~\ref{fig_comp}.

As already shown in Figure~\ref{fig_histo}, the mass distribution for 
the PMS  stars in our Tr\,14 sample peaks at $\sim 1$\,M$_{\odot}$ and
drops at lower masses because of photometric incompleteness  (see also
solid line in Figure~\ref{fig_comp}). On the other hand, the catalogue
of B11 samples a significantly different range of masses in
Tr\,37. The histogram (dashed line in Figure~\ref{fig_comp}) clearly
indicates that the majority of PMS objects in this star forming region
have masses from $\sim 0.1$\,M$_{\odot}$ to $\sim 0.6$\,M$_{\odot}$,
with a peak at $\sim 0.3$\,M$_{\odot}$. Since
the mass accretion rate $\dot{M}_{acc}$ scales
approximately linearly with the mass of the objects
\citep[e.g.][]{de11a, ba11, spe12, de13a}, the large difference in the
sampled masses in the two datasets is a plausible explanation for the
difference in the observed distribution of $\dot{M}_{acc}$.

%

\section{Spectral energy distributions}

By combining the WFI optical photometry with the SOFI and Spitzer/IRAC
datasets described in Section~\ref{sec_obs}, we are able to derive the 
spectral energy distributions (SEDs) of the stars in the central  $\sim
5\arcmin$ around the cluster centre (see Figure~\ref{fig_fov}). The
flux at each wavelength has been corrected for extinction, according
to the reddening values derived in Section~\ref{sec_redd} for the
individual objects.

\begin{figure*} 
\centering 
\includegraphics[width=0.9\textwidth]{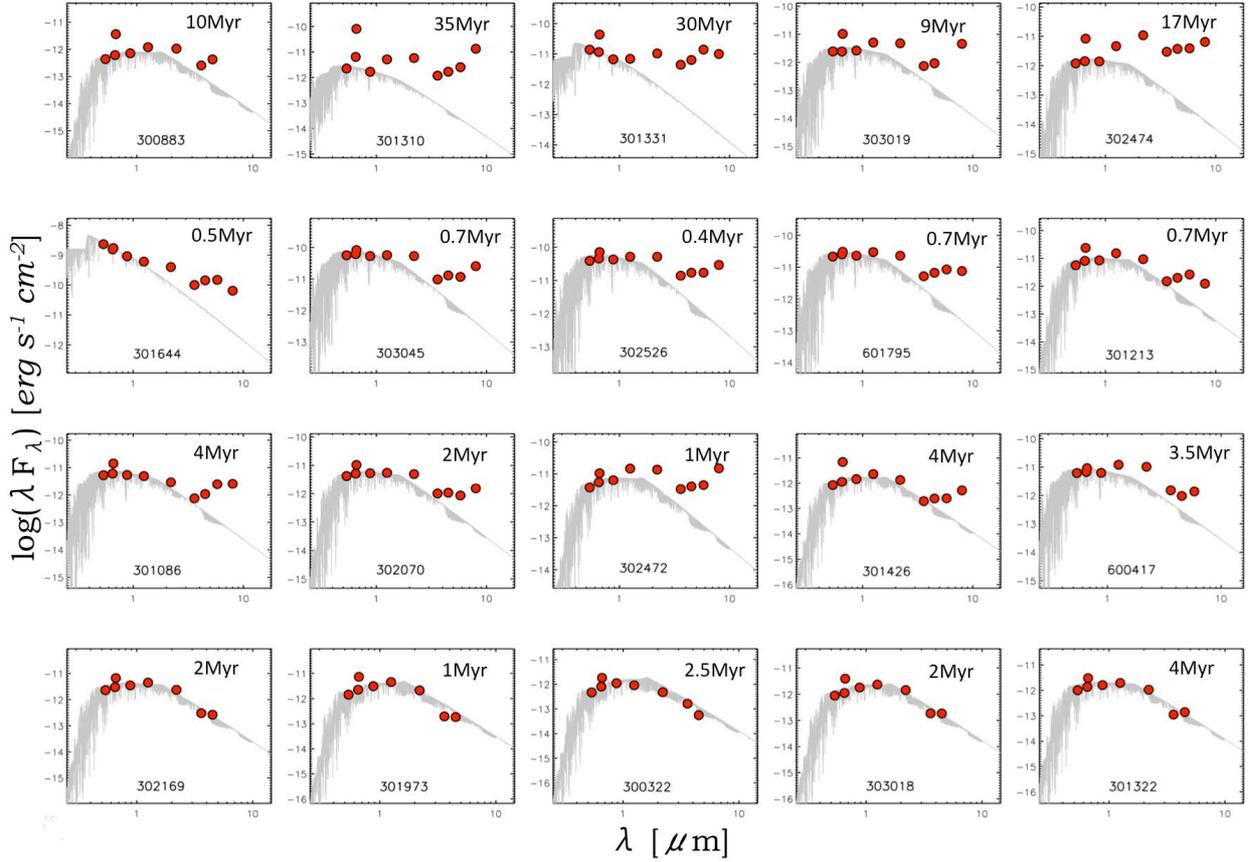}
\caption{SEDs of a sample of 20 bona-fide PMS objects labelled with their IDs as per
the photometric catalogue. The age of each object is also given. The
extinction corrected fluxes are represented  by the thick dots. The
error bars are smaller than the symbols and as such not shown. In grey
we show the NextGen stellar models that correspond to the physical
parameter of each star, as derived in Section\,5. }
\label{fig_sed}
\end{figure*}

Of the 389 bona-fide PMS stars, 57 fall outside the area covered by 
the other photometric catalogues, while 227 PMS have at least a SOFI and/or
Spitzer counterpart. We find these numbers quite reasonable, given that
both the SOFI and Spitzer observations are considerably shallower than
the WFI images and do not allow one to detect
the lowest mass PMS stars observed in the optical bands. Moreover, the lower
angular resolution of the near-IR observations further reduces the detection limits
of the Spitzer and SOFI data.
Nevertheless, the available data allow us for the
first time to collect a sample of 227 SEDs of  bona-fide PMS stars in
Tr\,14, some of which are shown in Figure~\ref{fig_sed}. We note that
the error bars in the figure are smaller than the symbol size due to 
the low flux uncertainties, and therefore are not shown. 

Over-plotted on the observed SED of each star (thick dots), we show a 
synthetic spectrum (grey line) from the NextGen stellar atmosphere 
models~\citep[][]{al12} for
the specific stellar parameters that we have derived in
Section~\ref{sec_pms}. We stress here that the models are not fitted to
the observed SEDs but are directly calculated for the measured stellar
parameters from our  photometry.

The five SEDs in the top row of Figure~\ref{fig_sed} are for older PMS
objects, while those shown in the second and third row are for PMS 
stars with ages younger than 5\,Myr. Finally, the bottom row shows the
SEDs for five bona-fide PMS stars that have excess H$\alpha$ emission
but no IR excess at the wavelengths covered by these observations.  A
visual inspection reveals that there is an excellent agreement between
the atmospheric models and the observed SEDs. 

It has been convincingly shown that the SEDs can be
efficiently used to detect discs around PMS objects and to study their
nature and evolution~\citep[see e.g.][and references
therein]{es10,fa13,me10}. Moreover, the slope of the SED in the
wavelength region covered by the near-IR excess, i.e. the spectral index
$\alpha$,  can be used to characterize different object
classes~\citep[][; see also  Alcal{\'a} et al. 2008 and references
therein]{la06}. However, we stress that in the present work we are
mainly interested to use the SEDs in order to identify stars with
near-IR excess. Among the 227 PMS with an IR detection (either SOFI,
Spitzer, or both),  82  (i.e. $\sim 36\,\%$) show near-IR excess at the
sampled wavelengths. Only $\sim 10\,\%$ of them have ages older than
10\,Myr, in relatively good agreement with the predictions made so far
on the lifetime of accreting protoplanetary discs and based mainly on
spectroscopic studies of nearby star-forming regions~\citep[e.g.][]{fe10}. 
Finally, it's interestingly to note that $\sim40\%$ and$\sim$47\% of the 
young and old PMS objects respectively, show IR excess i.e., show a 
disk at the sampled IR wavelength.

At first sight, the number of H$\alpha$ excess emitters detected in our
optical images that also show an excess in the SOFI and Spitzer images
might seem low. It is important to recall, however, that the observed 
SEDs in the near-IR depend on the opacity of the disc, which is governed
by the dust. In particular, dust opacity is affected by the composition
of the dust and by grain growth and settling. For example \citet[][]{es12}
showed that, for the same dust-to-gas mass ratio, discs with small
grain  size are more flared than discs with large ones. However, the
detection limit of the SOFI and Spitzer images and the lack of
observations at wavelengths longer than 10\,$\mu$m do not allow us to
investigate in detail the nature and the physical state of the discs in
most of the bona-fide PMS stars detected in this work.

It is possible that most of the older PMS stars in our sample should be classified
as transitional phase characterized by dust removal in the inner 
regions resulting in a deficit of infrared flux.
This deficit has become the defining characteristic of transitional disks~\citep[TDs][]{me10}. 
Our results would imply a scenario in which a
transitional disc still hosts a significant amount of gas accreting onto
the star, even at ages older than 10 Myr. Similar results were already obtained by~\citet[][]{zu95} but interestingly,~\citet[][]{ma14}
recently published a spectroscopic study of the gas content, accretion and wind properties of a sample of 22 
TDs with X-Shooter. They find that 80\% of the TDs in their sample show $\dot{M}_{acc}$ compatible with 
those of classical TTauri obects. Moreover, they conclude that in their somehow limited
sample of TDs there is a gas rich inner disk with density similar to that of disks in classic PMSs.
~\citet[][]{ber13}, using observations from the Herschel Space Observatory PACS
Spectrometer, found that the old ($\sim10$\,Myr) PMS star TW\,Hydrae is
surrounded by a gas disc with mass exceeding $0.05$ M$\odot$. This
surprising result indicates that TW\,Hydrae is an old PMS star containing a 
massive gas disc that is several times the minimum mass required to make
all the planets in our solar system. Even more recently,
\citet[][]{rob13} and~\citet[][]{mo13} reported the discoveries of a 
low-mass protoplanetary-like gaseous disc, together with debris discs,
around the older PMS stars 49\,Ceti and HD\,95086, with ages ranging
from 10 to 30\,Myr. All these results support a scenario in which
old and evolved debris discs still host a significant amount of gas and
confirm our results with completely independent measurements.

\section{Summary}

In this work we present optical observations of PMS stars in the
cluster  Tr\,14 in the Carina Nebula, obtained with the WFI imager at La
Silla. We further combine  these optical observations with near-IR
observations made with SOFI and Spitzer, in order to provide a more
complete description of the cluster's young stellar populations. The
main results can be summarized as follows:

\begin{enumerate}

\item 
We present the deepest optical CMD to date for this cluster.

\item 
Using H$\alpha$ photometry we are able to identify 356 bona-fide PMS
stars showing H$\alpha$ excess emission with EW$(H\alpha)>20\,\AA$. We
interpret this observational feature as indication that these objects
are still actively accreting gas from a circum-stellar disc. 

\item 
From the comparison of the HR diagram
with PMS evolutionary models we derive ages and masses of the 
PMS stars. We find that most of the PMS objects are younger than
10\,Myr with a median age of $\sim 2$\,Myr. Surprisingly, we find that
$\sim 20\,\%$ of the mass accreting objects are older than 8\,Myr. 

\item
The presence of candidate long-living accreting discs is in line with
the findings in a number of Galactic~\citep[][]{be10, de13b} and
extragalactic~\citep[][]{pa00, de10,de11a,de11b,spe12,de13a,de13b} 
starburst clusters.

\item  

Using the derived stellar physical parameters (mass, radius) of the PMS
objects and their H$\alpha$ luminosities $L(H\alpha)$, we derive the
accretion luminosity $L_{acc}$ and mass accretion rates $\dot{M}_{acc}$.
We find that $\dot{M}_{acc}$ decreases with time, in line with the
predictions of  models of viscous discs, but the $\dot{M}_{acc}$ that we
measure are systematically higher than those predictions. Taking into account
the photometric completeness in the H$\alpha$ observations  and our
requirement that a bona-fide PMS star be classified as such only if it
has EW$(H\alpha) > 20\,\AA$, we determine the detection limit to the
measured $\dot{M}_{acc}$ rates. For comparison, we apply the same
analysis to PMS stars in Tr\,37, where the  distribution of PMS masses
peaks at $\sim 0.3\,$M$_{\odot}$. We demonstrate that our
$\dot{M}_{acc}$ values are reliable and that it is the mass distribution of
the PMS objects sampled in Tr\,14 that affects the measured $\dot{M}_{acc}$.
Indeed, a K--S test demonstrates that, once  the same mass range is
selected, the $\dot{M}_{acc}$ distributions in Tr\,14 and Tr\,37 are 
indistinguishable from one another.  

\item 

We finally combine the optical WFI photometry with near-IR observations
from SOFI and Spitzer, all corrected for extinction, in order to build
SEDs for each PMS star in Tr\,14. A comparison with synthetic spectra,
calculated using NextGen atmosphere models for the specific physical
parameters measured for these stars, indicates that only a small fraction
of them have near-IR excess, at any age. The lack of data at wavelengths
above $10\,\mu$m does not allow us to investigate in detail the nature
and the physical properties of the discs. However, our analysis suggests  the
presence of transitional discs in which a large amount of gas is still
present and sustains accretion onto the PMS object at ages older  than
10\,Myr. This scenario is supported by the recent discoveries with
Herschel of transitional discs containing a massive gas component around
the relatively old PMS stars TW\,Hydrae, 49\,Ceti, and HD\,95086. Our
results are in full agreement with these findings and together with them
they challenge the common-wisdom understanding of  circumstellar disc
evolution, possibly implying a new scenario for the planet formation
mechanism. The existence of a population of old discs suggests that
the planet formation process could proceed on much longer timescales
than previously thought.

\end{enumerate}

\begin{acknowledgements}

We wish to thank the anonymous referee for insightful comments
that have helped to improve the presentation of our work. This research has made use of the NASA/IPAC 
Infrared Science Archive, which is operated by the Jet Propulsion Laboratory, 
California Institute of Technology, under contract with the National Aeronautics and Space Administration.
NP acknowledges partial sup- port by STScIÐDDRF grant D0001.82435.
MZ was supported by Proyecto Fondecyt Regular 1110393, The BASAL Center for Astrophysics 
and Associated Technologies PFB-06, and from the Ministry of Economy, Development, and 
Tourism's Iniciativa Cient'fica Milenio through grant IC12009, awarded to the Millennium Institute of Astrophysics MAS.

\end{acknowledgements}

\end{document}